\documentclass[aps,prl,twocolumn,
showpacs,preprintnumbers,amsmath,amssym,superscriptaddress]{revtex4}
\usepackage{graphicx}
\usepackage{bm}

\begin{document} 

\title{
Integration of Langevin Equations with Multiplicative Noise
and the Viability of Field Theories for Absorbing Phase Transitions}

\author{Ivan Dornic}
\affiliation{CEA -- Service de Physique de l'\'Etat Condens\'e,~CEN~Saclay,~91191~Gif-sur-Yvette,~France}

\author{Hugues Chat\'e}
\affiliation{CEA -- Service de Physique de l'\'Etat Condens\'e,~CEN~Saclay,~91191~Gif-sur-Yvette,~France}

\author{Miguel A. Mu\~noz}
\affiliation{Instituto~de~F\'\i sica~Te\'orica~y~Computacional~Carlos~I,~Facultad~de~Ciencias,~Universidad~de~Granada,~18071~Granada,~Spain}

\date{\today}

\begin{abstract}
Efficient and
accurate integration of stochastic (partial) differential equations
with multiplicative noise can be obtained through a split-step scheme,
which separates the integration
of the deterministic part from that of the stochastic part, the latter
being performed by sampling
exactly the solution of the associated Fokker-Planck equation.
We demonstrate the  computational power of this method by applying it to most absorbing phase
transitions for which Langevin equations have been proposed.
This provides precise estimates of the associated scaling exponents,
clarifying the classification  of these nonequilibrium problems,
and confirms or refutes some existing theories.
\end{abstract}

\pacs{05.70.Ln, 
05.50.+q,
02.50.-r,
64.60.Ht
}
\maketitle

Stochastic differential equations are ubiquitous in the description
of phenomena in the natural sciences and beyond 
\cite{Gardiner}.
The coarse-graining of fast degrees of freedom often leads to 
effective Langevin equations where the noise term involves the mesoscopic 
variables of interest in a {\it multiplicative} fashion.
Examples range from nonlinear quantum optics, synchronization 
of oscillators, or wetting phenomena, to theoretical population 
dynamics studies and autocatalytic chemical reactions 
\cite{Gardiner,Review_MN,Review_Haye}.

An important case in nonequilibrium statistical physics
is the stochastic partial differential equation governing a single,
positive, concentration field $\rho=\rho({\bf r},t)$:
\begin{equation}
\label{sde_DP}
{\partial_t}\rho({\bf r},t)=D\, \nabla^2\rho +a \rho-b \rho^2
+\sigma \sqrt{\rho}\,\eta({\bf r},t),
\end{equation}
where  $\eta$
is a Gaussian (zero-mean) white noise (that is
 with correlations
$\langle \eta({\bf r},t) \eta({\bf r}',t') \rangle=\delta({\bf r}-{\bf
r}')\delta(t-t')$). 
For instance, for the reaction-diffusion process $A \to 2 A$, 
$2A \to 0$, Eq.~(\ref{sde_DP}) can be obtained in a variety of ways,
either from phenomelogical
considerations or through more rigorous transformations \cite{Mapping}.
Also named ``Reggeon field theory'' for historical reasons,
Eq.~(\ref{sde_DP}) describes the most prominent class of
absorbing phase transitions (APT), the directed percolation (DP) class 
\cite{Review_Haye}.
Indeed, interpreted
in the It\^o (prepoint) sense,
the unique, homogeneous $\rho=0$ solution does not evolve: it is an 
{\it absorbing state}. 
Although a wealth of models have been found to exhibit a DP transition,
this class does not encompass all possible cases,
and the classification of APTs is currently a very active field 
\cite{Review_Haye,Review_Odor}. Not only such an endeavor
is of importance for conceptual reasons,
but it should also yield a better understanding of the key ingredients
which have impeded so far clear-cut experimental realizations of even
the DP transition.

Following this line of thought, 
stochastic equations similar to (\ref{sde_DP}) have been proposed as 
candidate field theories for related problems (see below). Their analyses
are notoriously difficult, and mostly rely on the perturbative
renormalization group machinery in the vicinity
of the corresponding upper critical dimension, one of the few exceptions
being a recent 
non-perturbative treatment of Eq.~(\ref{sde_DP}) 
in \cite{delamotte_nprg_prl03}.
Given this analytical bottleneck, it is tempting,
 with ever-improving numerical 
resources, to directly integrate such stochastic equations in order to check 
whether they at least exhibit the universal properties they are supposed
 to represent.
However,  standard schemes  
either immediately run into severe difficulties.
For instance, even for the zero-dimensional version of Eq.~(\ref{sde_DP}), 
a first-order explicit Euler method, viz.
$ \rho(t+\Delta t)=\rho(t)+\Delta t[a \rho(t) -b\rho^2(t)]
+\sigma \sqrt{\Delta t\rho(t)}N(0,1)$,
where $N(0,1)$  is a normal random variate,
 will ineluctably 
 produce unphysical negative  values for $\rho(t+\Delta t)$,
and all the more so when $\rho \to 0$, the regime of interest for the APT.
Another route, which would first trade the 
square-root noise for a less singular one  
through some change of   variables (e.g., $\rho \to \psi^2$, or
a Cole-Hopf transformation $\rho \to e^{\pm 2 \phi}$),
 is also numerically unbearable 
since it  generates  pathological 
deterministic
terms as the original variable $\rho \to 0$.
Faced with this problem in the same context,
 Dickman proposed \cite{Dickman_scheme},
somewhat ironically, to also 
discretize 
 $\rho$,
yielding a scheme consistent  to the order ${\cal{O}}(\sqrt{\Delta t})$
in the limit $\Delta t \to 0$.
 This approach has been used with some success 
\cite{Dickman_scheme,Munoz_pdemany97,Munozetal_fes_num}, 
but one can legitimately wonder to what extent one is 
truly simulating the original, continuous equation.
In this respect, that the 
associated results are affected by the same long transients as those 
observed in microscopic models is also worrisome.

In this Letter, elaborating upon a method pioneered
by Pechenik and Levine in the somewhat distant context of 
front selection mechanisms in microscopic reaction-diffusion models
\cite{PL}, we  overcome the above hurdles. 
We first demonstrate the power of this approach
on Eq.~(\ref{sde_DP}) before applying it to most related APTs 
for which a Langevin equation has been proposed, including
the voter critical point with its two symmetric absorbing states
\cite{Voter_dickman,Voter_us}.
Our results are particularly worthy in the context of the current debate 
about APTs occurring when $\rho$ is coupled to an auxiliary field $\psi$:
when $\psi$ is static and conserved 
(Manna sandpile model, conserved-DP, or fixed energy sandpiles class) 
\cite{FvW,Alexv_PRL,Alexv_PRE,Fes_PRL,Munozetal_fes_num,JK_HC_PRL2}
we obtain the best numerical estimates for the critical indices.
When $\psi$ is conserved but diffuses
\cite{KSS,WOH,WOH_JSP,Freitas00,Janssen01,Freitas01,Janssen04},
our results suggest that,  at least in low spatial dimensions,
 the Langevin equations  postulated or  derived (approximately) 
 as candidate field theories  are  not viable.

The idea underlying the 
approach of \cite{PL}
(of which we become aware while this paper was upon completion)
is a general and rather natural one,
since it consists in integrating the fast degrees of freedom.
An It\^o stochastic differential equation of the form 
$\frac{{\rm d}\rho}{{\rm d}t}= f(\rho)+\sigma g(\rho) \eta(t)$
is dealt with the so-called operator-splitting scheme: the
 stochastic part $\sigma g(\rho) \eta(t)$ is integrated first,
{\it not} by using a Gaussian random number, but by directly 
sampling the time-dependent solution of the associated 
Fokker-Planck equation (FPE). Namely, 
one generates a random number $\rho^*$ 
distributed according to the conditional transition probability 
density function (p.d.f.)
${\rm Proba.}\{\rho(t+\Delta t)=\rho^*|\rho(t)=\rho_0\}$,
and then use $\rho^*$ for evolving the deterministic part $f(\rho)$
 with any standard numerical method for ordinary differential equations.
Since the integration of the stochastic part is accomplished
through the exact solution of the FPE,
which is  first-order in time,  the   overall precision of
 the  scheme,  ${\cal{O}}(\Delta t)$,
is already significantly superior to that of a naive 
 Euler method (anyhow flawed for Eq.(\ref{sde_DP})), 
or to Dickman's approach.

Now, for the square-root noise case, i.e., $g(\rho) \equiv \sqrt{\rho}$,
the closed form
solution  $P(\rho,t)={\rm Proba.}\{\rho(t)=\rho|\rho(0)=\rho_0\}$ 
of the associated FPE
$\partial_t P(\rho,t) =\frac{\sigma^2}{2}\partial^2_\rho [\rho P(\rho,t)]$,
 has been known 
in the mathematical literature for more
than half a century \cite{Feller} (see also \cite{PL}):
\begin{equation} 
\label{Bessel1} P(\rho,t)=\delta(\rho) e^{-\frac{2 \rho_0}{\sigma^2 t}}+ \frac{2 e^{-\frac{2(\rho_0+\rho)}{\sigma^2 t}}}{\sigma^2 t} \sqrt{\frac{\rho_0}{\rho}} I_{1}\left(\frac{4 \sqrt{\rho_0 \rho}}{\sigma^2 t}\right),
\end{equation}
($I_1$ is the modified Bessel function of the first kind of order 1).
When, further, the deterministic part is linear, 
i.e. $f(\rho)=\alpha+\beta \rho$, with $\alpha >0$, 
the exact conditional transition p.d.f. of the full equation 
$\frac{{\rm d}\rho}{{\rm d}t}= \alpha+\beta \rho+\sigma \sqrt{\rho} \eta$ 
has also been determined \cite{Feller}:
\begin{equation}
\label{Bessel2}
P(\rho,t)=\lambda e^{-\lambda (\rho_0 e^{\beta t}+\rho)}
\left[\frac{\rho}{\rho_0 e^{\beta t}}\right]^{\frac{\mu}{2}}
I_{\mu}\left( 2 \lambda \sqrt{ \rho_0 \rho e^{\beta t}}\right),
\end{equation}
($I_{\mu}$ being a Bessel function of order $\mu$)
 where, to condense notations, we have set 
$\lambda= \frac{2 \beta}{\sigma^2 (e^{\beta t}-1)}$, and
$\mu=-1+\frac{2 \alpha}{\sigma^2}$.

The scheme we have used to integrate Eq.(\ref{sde_DP}) and its siblings
relies on the latter results.
After discretizing the Laplacian $\nabla^2 \rho$
over the $2d$ nearest-neighbors ${\bf r}+{\bf e}_v$ of 
site ${\bf r}$ on a $d$-dimensional hypercubic lattice of mesh size
$\Delta x$, we first sample, between $t$ and $t+\Delta t$, 
the solution of the FPE associated to
each local linear equation
$\frac{{\rm d}\rho}{{\rm d}t}= \alpha+\beta \rho+\sigma \sqrt{\rho} \eta$ 
using Eq.(\ref{mix_num}) with $\beta=a- \frac{2 d D}{(\Delta x)^2}$ and
\begin{equation}
\label{eq_alpha}
\alpha =\alpha({\bf r},t)=
 \frac{D}{(\Delta x)^2}\sum_{v=1}^{2 d}\rho({\bf r}+{\bf e}_v, t).
\end{equation}
The value
 $\rho^*$ coming from the  stochastic sampling step is, by construction,
 automatically non-negative, and serves
as the initial condition for the remaining part of Eq.(\ref{sde_DP}), i.e.
 $\partial_t \rho({\bf r},t)=- b \rho^2({\bf r},t)$, 
which can be trivially integrated to yield
$\rho({\bf r},t+\Delta t)=\frac{\rho^*}{1+\rho^* b \Delta t}$.
Given that $b>0$,   the non-negativity of $\rho({\bf r},t)$ 
will be preserved at all times 
 if, initially, $\rho({\bf r}) \ge 0$ everywhere,
since $\alpha$ given by Eq.(\ref{eq_alpha}) will also
be non-negative and  Eq.(\ref{mix_num}) can be used.

It remains to sample the above p.d.f.,
 Eq.(\ref{Bessel1}) or Eq.(\ref{Bessel2}). 
Instead of using a table method, as the authors of \cite{PL},
we remark that, with the help of the Taylor-series expansion of
 the Bessel function, Eq.(\ref{Bessel2}) can be rewritten as
\begin{equation}
\label{mixture}
P(\rho,t)=\sum_{n=0}^{\infty} 
\frac{\left(\lambda \rho_0 e^{\beta t}\right)^n e^{-\lambda \rho_0 e^{\beta t}}}{n!}\frac{\lambda e^{-\lambda \rho} 
(\lambda \rho)^{n+\mu}}{\Gamma(n+\mu+1)}.
\end{equation}
In other words,  one has the following mixture  \cite{Devroye}:
\begin{equation}
\label{mix_num}
\rho^* ={\rm Gamma}[\mu+1+{\rm Poisson}[\lambda \rho_0 e^{\beta t}]]/\lambda,
\end{equation}
where 
${\rm Prob.}\{{\rm Poisson}[\lambda \rho_0 e^{\beta t}]=n\}= 
\frac{\left(\lambda \rho_0 e^{\beta t}\right)^n e^{-\lambda \rho_0 e^{\beta t}}}{n!}$, 
and  
${\rm Prob.}\{{\rm Gamma}[\omega]=v\}= \frac{e^{-v}v^{\omega -1}}{\Gamma(\omega)}$.
This procedure  will reconstitute, on average, all the terms of 
Eq.(\ref{mixture}) with their correct probability, and gives us
--- since  standard and uniformly fast
generators of Poisson and Gamma random numbers are
available  ---  a means of sampling 
in a ``numerically exact'' way these p.d.f. \cite{NOTE}.

\begin{figure}[ht]
\includegraphics[width=8.6cm,clip]{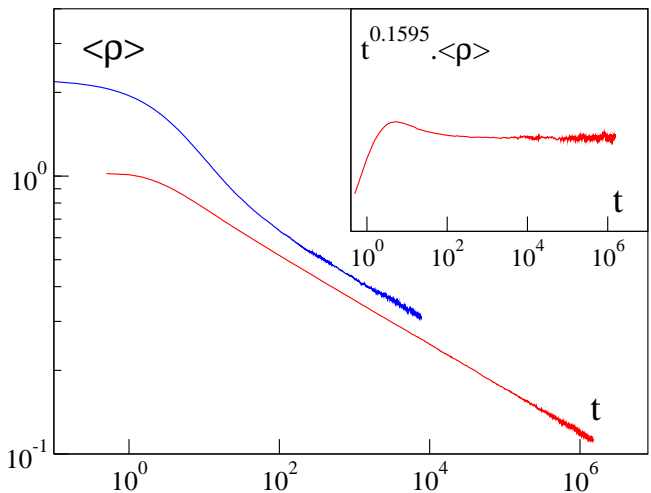}
\caption{(Color online)
Density decay $\langle \rho \rangle
 = \langle \rho({\bf r},t) \rangle_{\bf r} \sim t^{-\theta}$
at criticality for  Eq.(\protect{\ref{sde_DP}}) in $d=1$.
Lower curve: using our scheme ($\Delta t=0.25$, $\Delta x=1$,
$a=a_{\rm c}=1.75623(2)$,  $b=1$, $D=0.25$, $\sigma^2=2$,
 single run for a system of $2^{22}$ sites with
$\rho=1$ everywhere initially); 
a least-square fit gives $\theta=0.1595(2)$.
Upper curve:
using Dickman's method (similar conditions, but $\Delta t=10^{-3}$).
Inset: Plateau of the local exponent.}
\label{figure_1}
\end{figure}

Typical results for Eq.(\ref{sde_DP}) in one dimension 
are shown in Fig.~\ref{figure_1},
along with data  obtained using Dickman's
method. 
 Except for the (weak)  linear stability requirement
coming from the discretized Laplacian, there is no limitation on 
$\Delta t$ with the former method, so that the computational gain is 
of several orders of magnitude, together with 
an unusually clean algebraic decay of $\langle \rho \rangle$, with an
exponent  $\theta = 0.1595(2)$ 
matching to the fourth decimal  the  series-expansion result
\cite{Review_Haye,Review_Odor}.
In fact, even if $\Delta t=0.25$ for this run,
 the threshold $a=a_c(\Delta t)$  is  within one percent off its 
extrapolated limit value as $\Delta t\to 0$, suggesting that the 
 continuous limit of Eq.(\ref{sde_DP}) is already resolved.
One of  the reasons for the particular efficiency of this scheme
even with such a large timestep is that it  automatically takes
into account,
and in a self-adaptive fashion through the locally varying value of 
$\alpha$ (Eq.(\ref{eq_alpha})), the strongly 
non-Gaussian modifications undergone by  the
instantaneous, conditional p.d.f., Eq.(\ref{Bessel1}) or
Eq.(\ref{Bessel2}), as one gets closer and closer to the absorbing barrier.

We now present some of our most salient results  obtained
for Langevin equations similar to (\ref{sde_DP}),
 deferring a  more detailed 
account of our investigations  to \cite{TBP}.

\begin{figure}[ht]
\includegraphics[width=8.6cm,clip]{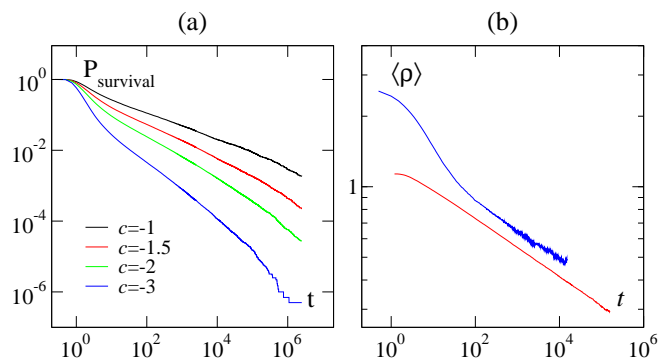}
\caption{(Color online) (a): Survival probability of an initial seed
at ${\bf r}={\bf 0}$
 evolving under  Eq.(\protect{\ref{sde_DP}}) with the extra 
term $c \exp[-w \int_0^t {\rm d}s\,\rho({\bf r},s)]$. 
Same parameters as in
Fig.~\protect\ref{figure_1} and various $c$ values (about $10^7$ trials).
(b): Same as Fig.~\protect\ref{figure_1} but for 
Eqs.(\protect{\ref{sde_Manna}}).
Lower curve: with our scheme ($\Delta t=0.1$, $\Delta x=1$,
$a=a_{\rm c}=0.86455(5)$,  $b=\omega=1$, $D=D_\rho=0.25$, $\sigma^2=2$,
$\rho=\phi=1$ everywhere initially);
A least-square fit gives $\theta=0.124(1)$.
Upper curve: with Dickman's method ($\Delta t=0.0025$).}
\label{figure_2}
\end{figure}

{\it  DP coupled to a non-conserved, non-diffusing field.}
To account for reaction-diffusion processes such as $2A\to 3A$, $2A\to A$ 
where single particles do {\it not} move 
(such as the prototypical pair-contact
process \cite{PCP}) and which thus possess infinitely-many absorbing states, 
it has been proposed that Eq.(\ref{sde_DP}) be
supplemented by the non-Markovian term 
$c \exp[-w \int_0^t {\rm d}s\,\rho({\bf r},s)]$
to account for the memory effect introduced by immobile particles
\cite{DP_manyabs}.
The impact of this term is however unclear, with early simulations 
\cite{Munoz_pdemany97}
using  Dickman's method suggesting that continuously-varying
 spreading exponents
arise, in agreement with results obtained on microscopic models
\cite{PCP}, but in contradiction
with the study of infinite-memory spreading processes \cite{inf_mem},
which support stretched-exponential behavior.
Simulations with our scheme, in one dimension, reveal power-laws
for small $|c|$, but curvature appears at late times for large, negative
$c$ values (Fig.~\ref{figure_2}a). 
To be fully conclusive, these results will have to be improved 
by using enrichment methods enabling to explore rare
events, but they already indicate that the conclusions of \cite{inf_mem}
probably hold asymptotically. 
We finally mention  that in two dimensions (and for $c >0$, $b=0$)  we obtain 
{\it dynamical percolation} spreading exponents as predicted by the 
standard theory \cite{DP_manyabs}.

{\it DP coupled to a conserved, non-diffusing field.} 
Reaction processes such as  $A+B\to 2A$, $A\to B$ where $A$ particles diffuse
and $B$ are static, are similar to the case above, but the number of 
particles is (locally) conserved \cite{Alexv_PRL}. This conservation law leads to couple
$\rho$ to a conserved field $\phi$ in the following 
system \cite{Alexv_PRE,Fes_PRL}:
\begin{equation}
\begin{array}{lll}
\partial_t \rho &=& D\, \nabla^2\rho +a\rho -b\rho^2 +\omega\rho\phi
 +\sigma \sqrt{\rho}\,\eta({\bf r},t)
 \\
\partial_t \phi &=& D_\rho\, \nabla^2\rho 
\end{array}
\label{sde_Manna}
\end{equation}
Microscopic models leading to (\ref{sde_Manna}) also include so-called
fixed energy sandpiles such as the Manna 
model, establishing a link between
APTs and self-organized criticality \cite{Fes_PRL}.
 The conservation law influences even
the static exponents but definite estimates are currently not available 
(see \cite{Munozetal_fes_num} and references therein). 
Data from microscopic models, as well as simulations of  (\ref{sde_Manna})
using Dickman's method are plagued by long transients/corrections
to scaling. 
Our scheme leads, again,
to clean power-laws which provide us with the best estimates
for the scaling exponents of this class of APT. In Fig.~\ref{figure_2}b,
we show a typical result for critical decay in $d=1$, 
leading to $\theta=0.125(2)$, unambiguously distinct from the DP value
0.1595(1). Critical decay exponents obtained in higher dimensions,
$\theta_{2d}=0.509(5)$ and $\theta_{3d}=0.81(1)$ \cite{TBP},
 differ     also     significantly from their DP counterparts.

{\it DP coupled to a conserved, diffusing field.} 
If, for the reaction processes above, both species are diffusing, the situation
changes again, if only because one has now a single, dynamic absorbing state
(where $B$ particles diffuse in the absence of $A$s). This case was studied
both analytically \cite{KSS,WOH,WOH_JSP,Janssen01,Janssen04} and numerically 
 \cite{Freitas00,Freitas01}
with continuous APT predicted and observed
for $0<D_\phi\le D$,
but with conflicting estimates of scaling exponents 
\cite{Janssen01,Freitas01}.
 The corresponding Langevin equation is usually cast 
\cite{WOH_JSP,Janssen04}
as Eqs.(\ref{sde_Manna}) complemented  by 
the self-diffusion of the auxiliary field  and a 
conserved noise term. 
Performing  with our scheme
critical decay experiments in spatial dimensions $d=1,2$
we find  the exponent $\theta$ to be undistinguishable
from  the DP values.
 Because this differs from
both analytical predictions \cite{KSS,WOH,Janssen01}
and estimates from microscopic
models \cite{Freitas01}, this indicates that the
truncation of the full action of the
field theory needed to arrive at the corresponding Langevin equations
is {\it not} legitimate.

\begin{figure}[ht]
\includegraphics[width=4.2cm,clip]{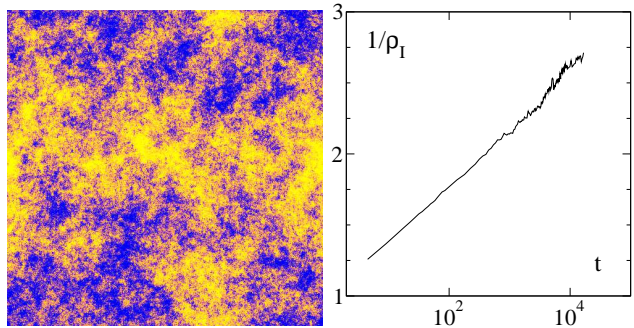}
\includegraphics[width=4cm,clip]{./fig3b.eps}
\caption{(Color online) Voter coarsening process for a continuous field
evolving under Eq.(\protect{\ref{sde_Voter}}) 
($\Delta x=1$, $\Delta t=0.25$, $D=1$, $\sigma^2=0.5$, zero-mean random
initial conditions)
Left: snapshot of $\rho$ at $t=2500$ ($512^2$ sites).
Right: decay of interface density ($4096^2$ sites)
$\rho_I=1-\langle\rho({\bf r},t)\rho({\bf r}+{\bf e}_v,t)\rangle
 \propto 1/\ln t$.}
\label{figure_3}
\end{figure}

{\it The voter critical point.} 
The universality class of the voter model is characterized by 
two symmetric absorbing states \cite{Voter_us}. 
The following field theory has been
proposed ---but never tested--- 
to describe its critical point \cite{Voter_dickman,Janssen04}:
\begin{equation}
\partial_t\rho = D\,\nabla^2\rho + \sigma\sqrt{1-\rho^2}\, \eta({\bf r},t)
\label{sde_Voter}
\end{equation}
The FPE associated to the sole stochastic part can be solved
through an eigenfunction expansion, leading to a complicated expression
for the conditional transition p.d.f., involving
a continuous part and two delta peaks at the barriers $\rho= \pm 1$
\cite{PL,TBP}.
Although this  distribution can be sampled \cite{TBP}, 
it is both much simpler and more efficient
to replace in  the noise term the piece $\sqrt{1-\rho^2}$
 by $\Theta(\rho)\sqrt{1-\rho}+ (\rho \leftrightarrow - \rho)$,
thereby taking into account just the  closest DP barrier.
 This way our scheme can be applied and, 
one observes in two dimensions 
phase ordering patterns typical of the marginal voter 
coarsening process, and, for the first time
with continuous variables, the expected $1/\ln t$ decay of the 
density of interfaces (Fig.~\ref{figure_3}).

This completes our (not-exhaustive) inspection of Langevin equations proposed
as field theories of absorbing phase transitions. 
Pending more comprehensive studies (higher dimensions,
other scaling exponents), the results already obtained demonstrate that
the method presented above  
enables faithful and efficient simulations of such stochastic
equations. This approach will remain particularly useful 
as long as no major analytical progress is made, and also to test 
future theoretical predictions.

M.~A.~M. acknowledges financial support from the Spanish MCyT (FEDER) under
project BFM2001-2841.


\begin{thebibliography}{99}
\bibitem{NOTE}
By continuity, 
Eq.(\ref{mix_num}) can also be used to sample 
  Eq.(\ref{Bessel1}), with the (natural) convention 
${\rm Prob.}\{{\rm Gamma}[0]=\delta(v)\}$.
\end{thebibliography}
\end{document}